\documentclass[12pt]{article}
\usepackage{slashed}
\def\bea{\begin{eqnarray}}
\def\eea{\end{eqnarray}}

\begin{document}
\title{Coulomb induced diffraction of energetic hadrons into jets}
\author{L.~Frankfurt, \\
Nuclear Physics Dept.,
        Tel Aviv University, Israel,\\
\\[8mm]
M.~Strikman,\\
Department of Physics, PSU, USA}

\maketitle
\begin{abstract}
The electromagnetic (e.m.) current conservation and 
 renormalizability of QCD are used to 
calculate 
the amplitude of energetic hadron(photon) diffraction
into several
jets with large relative transverse momenta off the  nucleon(nucleus)
Coulomb field. Numerical estimates of the ratio of e.m. and strong
amplitudes show  that 
within the kinematic range where
the leading twist 
approximation for the strong amplitude is applicable, 
the e.m.  contribution  can be neglected. In $pA$ scattering at LHC
and in the fragmentation of a photon into two jets in ultraperipheral $AA$
collisions in the black limit (which maybe realistic   at LHC) e.m. 
contribution
 may  win.
\end{abstract}
\maketitle 

\section{Introduction}

Diffraction of hadrons into jets off  the gluon field of a nucleon(nucleus) 
is becoming  an effective method of investigation of the  light cone wave 
functions of hadrons. The H1 and ZEUS  experiments 
at HERA  discovered processes of hard diffractive electroproduction 
of vector mesons off proton target \cite{AC}  with the striking properties
predicted in \cite{FMS93,Brod94,CFS}  based on  the new QCD factorization
theorems for high energy exclusive processes.  The  color transparency
phenomenon in the coherent pion dissociation into two high $p_t$ jets off
nuclear target has been predicted in  \cite{FMS93} and observed at FNAL 
\cite{ashery} including  a very strong ($\propto A^{1.55}$) dependence of the
cross section on the atomic number. Moreover the measured dependence of the  
differential cross section on  transverse and longitudinal components of the 
jet  momentum agrees 
well with the one which is characteristic for high momentum tail 
of pion wave function evaluated in  \cite{FMS2000}. A potential background for 
the hard QCD physics which has similar dependence on the atomic number 
$\propto Z^2/A^{1/3}$  arises from the inelastic diffraction of a  projectile 
pion off the Coulomb field of  the target. Similar background exists for the 
process of the photon  diffraction into two jets which is currently being
studied  by ZEUS(HERA)\cite{ashery1}. It is also present
 for the process of diffractive
proton fragmentation into three jets which is one of the promising new
hard processes for studies at RHIC, LHC and Tevatron \cite{FELIX}.  
Processes of high
energy hadron    scattering off the Coulomb field are interesting by
themselves  because their amplitudes are rapidly  increasing with 
the energy of the
collision as a result of the zero mass  of the photon and the  decrease with 
the incident energy of the minimum momentum transfer to the nuclear target,
$t_{min}$. At sufficiently high  energies these processes are
dominated by the scattering  at impact parameters significantly
exceeding the geometrical radius of a target. Therefore at sufficiently large
energies initial and final  state interactions should become negligible.
Thus there exists a practical necessity to evaluate legitimately hadron
diffraction into high $k_t$ jets in the ultraperipheral collisions. Moreover 
in the regime of color opacity (LHC, Tevatron?) proton dissociation into high 
$k_t$ jets from Coulomb field may dominate in  some kinematical region.
This is because in this regime the proton diffraction into three jets should
be strongly suppressed as compared 
to the  perturbative QCD (pQCD) expectations.

A nucleon (meson,photon) has an appreciable  amplitude to be in a state where 
valence partons are localized in a small transverse area. These configurations 
are usually referred to as {\it minimal} Fock space configurations -  
$\left|3q \right>$, $\left|q \bar q\right>$. A  significant amplitude of the 
decays $\pi(\rho)\rightarrow leptons$ as well as the observed significant 
cross  sections of the processes: $\gamma^*+p\rightarrow V+p$ for 
$V=\gamma,\rho,\omega,\phi,\psi,\psi'$ \cite{AC}, and of the process 
$\pi+A\rightarrow  2jet+A$ \cite{ashery} provide  the experimental evidences 
for the significant probability of such configurations in a photon, in  
pseudoscalar and vector mesons. The amplitude of the three quark configuration 
within proton can be estimated 
within the QCD inspired models, and in the long run it can be calculated in 
the lattice QCD. Note that the knowledge of this amplitude is important for 
the unambiguous calculation of the 
proton decay within the Grand Unification 
models. \footnote{Naively, there appears to be 
 a  contradiction between the experimental
observation that hadrons consist of an infinite number of partons and  the
dominance of minimal quark configurations  in the light cone hadron wave
function of hadrons required for describing hard diffractive processes. The 
QCD factorization theorem resolves this contradiction by including other 
partons into generalized parton distributions of nucleon(nucleus) target.}

In \cite{Gribov} V.Gribov  has derived a theorem for the radiation by 
electrically charged particles in two body collisions of an 
 energetic photon but 
with small momentum relative to the  scattering plane. The essence of the 
theorem is  that even at high energies the photon radiation is dominated by 
the photon emissions from external lines before or 
after the strong interaction.
 Later this theorem was used to
derive the QCD evolution equation in the Minkowski space  \cite{DDT}. In  
Ref. \cite{FMS2000} the Gribov technique was used to calculate the cross 
section of the process of pion diffraction  of Coulomb field of 
nucleon(nucleus): $\pi+A\rightarrow ~jet_1 + ~ jet_2 + A$.
It was demonstrated that 
the amplitude of the pion diffraction into jets off  the
Coulomb field of nucleus is calculable  unambiguously in terms of the high
momentum tail of the light cone pion wave function in 
the leading term in invariant energy, $s$,  
$\alpha_s$  and the transverse momentum of the jet, $k_t$.
 In this paper we
elaborate the analysis of \cite{FMS2000} and extend it to calculate 
the Coulomb 
dissociation of any hadron (proton, photon) into several jets corresponding 
either to a minimal Fock configuration, like three jets in the case of the 
proton, or the multi jet final states containing  a gluon jet which originate 
from the presence of $q\bar q g $ components in the meson wave functions.
We explain that amplitudes of the discussed processes can be unambiguously
calculated in QCD including the effects of the QCD evolution. First, we
derive the Williams-Weizecker approximation for the amplitude of 
ultraperipheral diffraction of a hadron into jets off nucleus Coulomb field 
and express the amplitude in terms of the field of equivalent transverse 
photons of the projectile hadron. We give  a more detailed explanation of
the key observation of \cite{FMS2000} that that the dominant contribution
into the 
equivalent photon field of projectile hadron is given by the interaction
of transversely polarized photon with the external quark lines within the
precision $O(\alpha_s(k_t))$. The physical interpretation of this result 
is that the radiation of an on-mass-shell photon with a small transverse 
momentum occurs after the strong interaction and therefore it does
 not distort the wave function of the hadron.

The relative importance of the strong interaction and e.m. mechanisms of
dissociation of protons and pions into jets is evaluated. The e.m.
mechanism is found to be a small correction in the 
energy and transverse momentum 
range where the leading twist effects dominate in the strong amplitude.
However if at very high energies the black body limit is reached 
at a certain range of $k_t$ the e.m. mechanism may become dominant.

\section{Evaluation of amplitudes}
\label{jets}

In this section we  evaluate the amplitude of a hadron $h$ diffraction 
into jets off the Coulomb field of a target of the atomic number $A$: 
$h+A\rightarrow jets +A$. At high energies the minimal momentum transfer
$-t_{min}R_A^2/3=
m_{N}^2\left({m_{jet}^2-m_h^2\over \nu}\right)^2 {R_A^2\over 3}\ll 1$, where
$\nu={2(p_h p_A)\over A}$, $R_A$ is the  e.m. radius of 
the target, and $m_{jet}^2$ is the invariant mass$^2$ of the produced
multijet state. Hence for $t \sim t_{min}$ which dominate in the total
cross section,  the process is  ultraperipheral and 
there are  no initial or final state  strong interactions  of $h$ or jets
with the target.

This amplitude is  due to the exchange of a virtual photon of four-momentum 
$q$ ($q^2=t$) with the target. The nuclear  Primakoff amplitude is then 
given by
\begin{equation}
{\cal M}_h (A) = e^2 {\langle h\vert J_\mu^{\rm em}
\vert jets \rangle\over -t}
  (P_A^i+P_A^f)^\mu {Z\over A}F_A(t)\approx {2 e^2}
 \langle h\vert J^{\rm em}\cdot {P_A\over A}
     \vert jets \rangle {ZF_A(t)\over -t},\label{prim}
\end{equation}
where $F_A(t)$ is the electric form factor of the nucleus.
Depending on the quantum numbers of a jet it can be described as a quark,
an antiquark, or a  gluon jet. A photon can be attached to any charged
particle, so a direct calculation of 
$\langle h\vert J^{\rm em}\cdot {P_A\over A}\vert jets\rangle$ 
involves a complicated sum of diagrams and would be 
rather complicated because of the cancellations between the
contributions of different diagrams. However we will considerably simplify
the calculation by using the conservation of the e.m.  current as 
well as the Sudakov variables. Accordingly, we write :
\begin{equation}
q=\alpha{P_A\over A}+\beta p_h+q_t.
\end{equation}
Conservation of the four-momentum gives
\begin{equation}
\beta={q^2\over 2(p_h\cdot P_A)},\quad\alpha
={m_h^2-m_{jet}^2-q^2\over \nu}
\approx {-m_{jet}^2\over \nu}.
\label{kin}
\end{equation}
Then the conservation of the e.m.current can be rewritten as
\begin{equation}  \langle h\vert  J^{\rm em} \cdot q \vert
 jets\rangle\approx
 \alpha \langle \vert  J^{\rm em}
\cdot{P_A\over A}\vert h\rangle
+\beta \langle h\vert  J^{\rm em}\cdot
p_h\vert jets\rangle + \langle h \vert
J^{\rm em}\cdot q_t\vert jets\rangle=0.
\label{cc}
\end{equation}
Using Eq.~(\ref{kin}) and keeping only the leading term in 
$\mu^2/\nu$,
where $\mu$ is the typical mass involved in the considered process, allows 
us to neglect the $\beta$ term in Eq.~(\ref{cc}) so that
\bea \alpha
\langle h\vert  J^{\rm em}\cdot
{P_A\over A}\vert jets \rangle= -
\langle h \vert  J^{\rm em}\cdot q_t\vert jets\rangle.
\label{wf}\eea
We want to stress that our formulae are accurate within the leading power
of $k_t$ only. 
So in the matrix element of the e.m.  current
$\langle h \vert
J^{\rm em}_{\mu}\vert jets\rangle$ we can safely put $q_t=0$.
By definition, the  transverse momentum of the  pion is zero, 
so the dominant (in powers of $k_t$) contribution in  Eq.~(\ref{wf}) is
given by the photon attachments to the external quark lines.
Hence in difference from the situation considered by V.Gribov there is no radiation from the initial state. The
matrix element is given by
\footnote{In \cite{FMS2000} the factor 2 in the numerator was missing.} 
\bea
\langle h \vert  J^{\rm em}\cdot q_t\vert jets \rangle=
\chi_h(z_i,k_{t,i}) \sum_{i} {2e_i (\vec{q_t}\cdot{\vec \kappa_{t,i}})\over 
z_i} 
\label{above}.\eea
Here $e_i$ is the electric charge of quark in the units of 
the electric charge of 
electron. The relative sign of each term can 
be easily visualized by  considering an 
antiquark as the quark moving in the backward direction. The factor
$1/z_i$ arises from the propagator of the 
interacting quark. This is because the  
propagator of the 
interacting quark multiplied by the factor $z_i$ should be
included into the definition of the light cone wave function of the
projectile hadron. The factor 2$k_{t,i}$ in the above  formulae arises as a 
consequence of the commutation of the operator of quark(antiquark) momentum 
$\slashed{k}_{i}$ from the quark propagator with 
$\slashed{q}_{t} $: 
$(\slashed{k} + \slashed{q})\slashed{q}_{t}=
2 (k_t\cdot q_t) - \slashed{q}_{t}\slashed{k}$, 
 and neglecting the last term. Indeed the operator $
\slashed{k}$ when
acting on the the spinor $u(k)$ leads to a term proportional to  $\sqrt{k^2}$, 
that is the virtuality of the final state (anti)quark which in the leading 
log approximation is $\ll k_t$ and hence can be neglected. Therefore 
the
effects related  to the virtuality of  quark(antiquark) in the final state
can be neglected 
in the leading order in 
 $\alpha_s \ln(k^2_t/\Lambda_{QCD}^2)$. 
The contribution of $\slashed{q}$ from the propagator of the 
interacting quark can be neglected as well since it is of the higher order 
in powers of $1/k_t$.  It is easy to check that in the case when quark
(antiquark) transverse momenta are ordered according to the DGLAP
approximation the equation derived above  coincides with the Gribov formulae
for the photon bremsstrahlung.

The generalization of this result to account for all Feynman diagrams having 
the same  powers of $\nu$ and $k_{t}$ is almost trivial. The  relative
contribution of other diagrams is $\propto k_{t}^{\prime}/k_{t}$ where 
$k_{t}^{\prime}$ is the transverse momentum of quarks in the intermediate 
state. 
Therefore the dominant contribution arises from the region of integration:
$k_{t}^{\prime~2} \simeq k_{t}^{2} $.
But within the $\alpha_{s} \ln k_{t}^2/\Lambda_{QCD}^2$  approximation
$k_{t}^{\prime ~~^2}\ll k_{t}^2$ so this contribution does not lead to
a $\ln k_{t}^2/\Lambda_{QCD}^2$ term. Thus the  above formula is valid
within the $\alpha_{s} \ln k_{t}^2/\Lambda_{QCD}^2$ approximation
when $\alpha_{s}\ll 1$.

Using Eqs.~(\ref{wf}),(\ref{prim}) we obtain the 
 final result:
\begin{equation}
{\cal M}_h (A) = \frac{-e^2 {\chi}_h (z_i,k_{t,i}) Z}
{q^2_t-t_{\rm min}}F_A(t){\nu \over m^2_f}
\sum_{i} {2e_i(q_t\cdot k_{t,i})\over 
z_i}\ d^{-n/2}(k_{it}^2).
\label{final}
\end{equation}
Note that in the
calculation of the cross section of photon (but not hadron) fragmentation
into jets one should multiply square of the above amplitude by the
 the number of colors.  
The wave function 
${\chi}_h (z_i, k_t)$ in the above equation describes 
 the high momentum tail of
the 
light cone wave function of a hadron defined as the equal time Bethe-Salpeter 
wave function which is normalized as in \cite{BL}. 
In the case of the gluon jets with transverse momenta $\ll$ quark transverse
momenta ${\chi}_h (z_i, k_t)$      is calculable in terms of the
 minimal Fock wave function
and formulae for the gluon bremsstrahlung.

In Eq.\ref{final} we wrote amplitude in
the form which is actually applicable for  
the
description of  the amplitude of the diffractive dissociation of any hadron 
$h$ off the Coulomb field into any number of jets including gluon jets. To
include gluon jets production one should substitute in the above formulae
the minimal Fock component of the hadron wave function by the component of
wave function containing gluons besides of valence quarks. Different
projectiles are characterized by the different  charges of quarks and by
the different number of valence quarks. Thus one should account that 
$m^2_f=\sum_{i}k_{it}^2/z_{i}$ is different for the system of two or three 
quarks. Above equation is the basic new result obtained in this paper. We 
want to draw attention to the curious feature of above expression - in the
case of projectile $\pi^{+}$ the amplitude of the process becomes zero
when $z_u=2/3$. This is due to the lack of the transverse dipole strength
for such  quark-antiquark configuration in $\pi^+$. Similarly for the
proton case the amplitude equal to zero when
${2(k_{t,u,1}\cdot q)\over z_{u,1}}
+ {2(k_{t,u,2}\cdot q)\over z_{u,2}}
- {(k_{t,d}\cdot q)\over z_{d}}=0$.

We want to draw attention 
that the renormalization of the wave function accounts for the cancellation 
of the infrared divergences \cite{BL}. This leads to the additional 
renormalization  factor $d(k_t^2)^{-n}$ in the cross section where n=2(3)
for a pion(proton) projectile. cf. \cite{FMS2000} (Here $d(k_t^2)$ is the
renormalization factor for the quark Green function). A rather straightforward 
method to avoid such infrared divergences is to use a special light cone
gauge in the calculation of  the matrix element of the e.m.
current:  $p_{h,\mu}A_{\mu}=0$\cite{FMS2000}. For the  proper definition
of jets one should sum also over collinear radiation. The high momentum tail
of the pion wave function has been calculated in \cite{FMS2000}.

The Primakoff term for the pion dissociation into two jets
has been evaluated also in the paper of D.Ivanov and
L.Szymanowski \cite{IS1}. Similar to \cite{FMS2000} they concluded 
that the e.m. contribution is  negligible. 
However the approximations made in \cite{IS1} and 
the results of  \cite{IS1} differ from ours. Our calculation is based on 
the generalization of the QCD factorization theorem which properly
accounts for the  conservation of the e.m. current and renormalizability of 
QCD. On the contrary \cite{IS1} restricted themselves by the set of diagrams 
where Coulomb photon may interact with an on-mass-shell q$\bar q$ pair in the
initial and final states only, which is in variance with the fact that the 
Coulomb interaction with a bound state is significantly more complicated. They
assumed that the q$\bar q$ pair in the intermediate states can be put 
on-mass-shell before the separation of scales and the cancellations between 
different photon attachments has been taken into account. These approximations
have problems with the conservation of the e.m.current and the renormalization 
group in QCD.

\section{Numerical estimates}

The ratio of e.m.  and strong amplitudes for the hadron diffraction
off a nucleus target into high $k_t$ jets is almost independent of the
projectile because most of the  factors are cancelled out in this ratio. 
In the case of the pion projectile this ratio 
has been evaluated in \cite{FMS2000} for $q_t^2=0.02 GeV^2, s=10^3 ~GeV^2 $,
$z=1/2$, $k_t=2 GeV$ and $A\sim 200$ to be about $ -i0.04$.  
Since typical $q_t$ are very small,  both the fixed target and collider
experiments can only measure the cross section integrated over $q_t$.
The strong amplitude is predominantly even under 
the transposition: $q_t\rightarrow -q_t$ while the  
e.m.  amplitude  is odd under  the transposition: 
$q_t\rightarrow -q_t$. Hence averaging over the angles the interference is 
canceled out.

The relative contribution of the square
of e.m.   amplitude to the total cross section
of  the jet production
can be estimated as 
$$R\approx {\vert M(Coulomb)\vert^2\over\vert M(strong)\vert^2} \bigg | _{q_t^2=0.02}*(0.02 B)\ln{1\over
-Bt_{min}},$$
where we used an exponential fit to the nuclear electric  form factor:
$F_{em}(t)=exp Bt$, 
where $B={R_A^2\over 6}$ and $R_A=A^{1/3} 1.1 Fm$. 

For the kinematics of the FNAL experiment we get for the ratio of 
e.m.  and strong contributions to the diffractive yield of jets
for heavy nuclei a value of about 
$5\cdot 10^{-3}$. For LHC two factors work in opposite directions -
the e.m.  contribution is enhanced by a factor of $k_t^2$ 
but suppressed due to a fast increase of the gluon density with decrease of 
$x$,  and increase of $k_t$. Taking, for example, 
 $s\approx 10^8 GeV$ and 
$k_t=10 GeV$  we find approximately the same ratio for e.m. 
and strong contributions.

For the case of proton-nucleus scattering we expect similar ratios,
though a more quantitative analysis would require a more detailed
 treatment of the strong amplitude.

In the estimates made in the paper we used the  energy
dependence of the pQCD amplitude as given by LO pQCD. At the same time 
there exists serious reasons to anticipate black body regime for the pQCD
physics in pA collisions at LHC, Tevatron  for jet transverse momenta
$k_t\le k_{t0}$ \cite{FELIX,fgms}. In this case pQCD amplitude will be 
strongly suppressed by the shadowing 
effects since 
  no inelastic diffraction is possible in the black body limit anywhere 
but near the nuclear edge. This leads to 
$\sigma \propto {A^{2/3}\over R_A^2}
\propto const(A)
$, 
while the A-dependence of the Coulomb contribution is not changed.
 So if the
interaction becomes black at some $k_t$ scale e.m.  contribution
starts to  win.

One practical possibility to investigate the e.m.  mechanism
and to measure the photon wave function would be to study 
at high-energy electron positron colliders  the  processes: 
$e^+ e^-\rightarrow e^+ +~two ~jets~ + e^-$,~\footnote{S.Brodsky,
 A comment at 
the Bad Honnef  workshop, 
July 2002} \, and  $e^+ e^-\rightarrow e^+ + ~three ~ jets~ + e^-$.

Formulae derived in the paper are equally applicable to the QED processes
like diffractive photoproduction of high $k_t$ lepton pairs off the nuclear 
field which has been considered firstly in \cite{bethe}, the diffractive 
fragmentation of positronium into high $k_t$ electron-positron pairs, etc. 
In these cases electric charges of constituents are equal but have opposite 
signs. As a result, the basic 
 equation  can be rewritten in terms of the derivative 
over $k_t$ of the light-cone wave function  of an electrically neutral 
projectile. This result has been anticipated in \cite{Hoyer}. In the case
of charged hadron fragmentation into  quark jets formulae have 
more complicated 
structure.

We are indebted to S.Brodsky who draw our attention to the paper of
Bethe and  Maximon
\cite{bethe} where single and double photon exchanges have been
calculated for the photoproduction of electron-positron pairs off
the nucleus Coulomb  field. We also thank  GIF and DOE for support.

\end{document}